\documentclass[12pt]{article}
\usepackage{a4,latexsym,amssymb,epsf}

\newcommand{\bm}[1]{\mbox{\boldmath $#1$}}
\newcommand{\starprod}[2]{\renewcommand{\arraystretch}{0}
                          \begin{array}{c}\scriptstyle #2\\ 
                          \mbox{\Huge $\ast$}\\ \scriptstyle #1 
                          \end{array}\renewcommand{\arraystretch}{1}}

\renewcommand{\theequation}{\arabic{section}.\arabic{equation}}

\begin{document}

\vspace*{4mm}
\centerline{\LARGE\bf Aperiodic Ising Quantum Chains}\vspace*{8mm}

\centerline{\large Joachim Hermisson${}^{\rm a}$, \
Uwe Grimm${}^{\rm b}$ \ and \ Michael Baake${}^{\rm a}$}\vspace*{8mm}

\hspace*{2cm}
\noindent${}^{\rm a}$Institut f\"{u}r Theoretische Physik,
Universit\"{a}t T\"{u}bingen, \hfill \break
\hspace*{2.8cm}
Auf der Morgenstelle 14, D--72076 T\"{u}bingen, Germany

\hspace*{2cm}
\noindent${}^{\rm b}$Institut f\"{u}r Physik,
Technische Universit\"{a}t Chemnitz, \hfill \break
\hspace*{2.8cm}
D--09107 Chemnitz, Germany 
\vspace*{8mm}

\begin{abstract}
Some years ago, Luck proposed a relevance criterion for the effect of 
aperiodic disorder on the critical behaviour of ferromagnetic Ising 
systems. In this article, we show how Luck's criterion can be derived 
within an exact renormalisation scheme for Ising quantum chains 
with coupling constants modulated according
to substitution rules. Luck's conjectures for this case are confirmed 
and refined. Among other outcomes, we give an 
exact formula for the correlation length critical exponent 
for arbitrary two-letter substitution sequences with marginal
fluctuations of the coupling constants.
\end{abstract}

\section{Introduction}

The influence of aperiodic (dis-)order on the thermodynamic properties
and the critical behaviour of spin systems has been an active
research area, particularly since the discovery of quasicrystals in 1984. 
In view of the many
articles existing, we cannot give proper credit to all contributors; 
instead, we would like to refer the reader to a recent review \cite{GB96}
which contains a rather complete bibliography on this subject.
For ferromagnetic Ising systems, 
heuristic scaling arguments, put forward by Luck \cite{L}, lead to 
qualitative predictions on the relevance of disorder implemented
through an aperiodic variation of the coupling constants which go beyond
the original relevance-irrelevance criterion of Harris \cite{Harris}.
For the Ising quantum chain (IQC), or the equivalent 2$d$ classical
Ising model \cite{Kogut} with layered disorder, this 
relevance criterion reads as follows: 
The critical behaviour is of Onsager-type
as long as the fluctuations in the sequence of coupling constants are bounded, 
whereas for unbounded fluctuations the critical singularities resemble those
of the randomly disordered case. This behaviour had first been conjectured
by Tracy \cite{Tracy} on ths basis of his results for 
two different examples of three-letter substitution rules. 
Of particular interest is the
marginal case with logarithmically (in the system size) 
diverging fluctuations. Here, 
the critical exponents are predicted to depend continuously on
the coupling constants \cite{L}.

Meanwhile, some exact results were obtained for 
substitution chains with marginal fluctuations of the couplings. 
For a number of substitution chains, the surface magnetisation 
critical exponent could be calculated \cite{TIB,BBHILMT,BBT,BB};
and, recently, the correlation length critical exponent was 
derived for two special substitution rules using an exact 
renormalisation approach \cite{IT}. 
All these results are in accordance with Luck's predictions.

The real space renormalisation method we use in this article is based
on the decimation process introduced in Ref.~\cite{IT}. Our setup is quite 
different, though; and in particular we are going to show how 
{\em arbitrary}\/ substitution rules can be analysed in terms of 
$S$-matrices and their star-product formalism.
In this way, the dependence of the critical behaviour on the nature 
of the fluctuations can be derived explicitly. The analysis is carried 
out in detail for two-letter substitution chains, with the result that
Luck's relevance criterion emerges from the renormalisation relations 
in a natural way. 
An extended renormalisation scheme for the most general case of $n$-letter 
substitution rules is presented in an Appendix and discussed for 
certain examples. For infinite classes of substitution 
chains with marginal fluctuations, including the 
two-letter chains, an analytic expression for the 
correlation length critical exponent is given.

This article is organised as follows. In Sec.~2, 
we give a brief outline of the  model, including some introductory 
remarks on substitution rules and on properties of the corresponding 
substitution sequences. Then, in Sec.~3, we introduce the 
renormalisation transformation for a restricted class of 
$n$-letter substitution chains, comprising all two-letter substitution chains. 
The dependence of the renormalisation flow on the fluctuations of the 
couplings is shown and the consequences are discussed in detail. 
Subsequently, in Sec.~4, we summarise the implications of our results
for the critical properties of the IQC for the three 
different types of fluctuations in the sequence of the coupling constants. 
Our conclusions and an outlook on future developments are presented
in Sec.~5. The paper contains two Appendices related to the content of
Sec.~3: In Appendix~A, the solution of the eigenvalue problem for
the case of a two-letter substitution sequence is derived; and finally, 
generalising the discussion of Sec.~3, the 
renormalisation transformation for arbitrary $n$-letter substitution 
rules is explained in Appendix~B.

\section{Ising quantum chain and substitution sequences}
\setcounter{equation}{0}

The IQC in a transverse magnetic field is defined by the Hamiltonian
\begin{equation}
H_{N}^{} = -\frac{1}{2}\left(\sum_{j=1}^{N} 
     \varepsilon_{j}^{}\,\sigma^{x}_{j}\sigma^{x}_{j+1} + 
     \sum_{j=1}^{N} h_j^{} \sigma^{z}_{j}\right)
\label{eq:QC}
\end{equation}
acting on the tensor-product space
$\bigotimes_{j=1}^{N}\mathbb{C}^{2}_{}\cong\mathbb{C}^{2^N}_{}$.
Here, the $h_j^{}$ denote the transverse fields at the sites. 
The $\varepsilon_{j}^{}$ are site-dependent coupling constants,
and the operators $\sigma^{x,z}_{j}$  
denote Pauli's matrices acting on the $j$\/th site. 
In what follows, we consider closed chains with
{\em periodic\/} boundary conditions which 
are defined by $\sigma^{x}_{N+1}=\sigma^{x}_{1}$; various other kinds of
boundary conditions can be chosen without changing the key results.

For a general set of coupling constants $\varepsilon_{j}^{}$ and fields
$h_j^{}$, the IQC can be written as a free-fermion model via a Jordan-Wigner
transformation and can then be diagonalised by a 
Bogoljubov-Valatin transformation, resulting in
\cite{LSM}
\begin{equation} \label{ffh}
H_N^{}=\sum^N_{q=1}\Lambda_q^{} (\eta_q^\dagger\eta_q^{} - 
{\textstyle\frac{1}{2})} + C
\end{equation}
where $\eta_q^\dagger$, $\eta_q^{}$ are $N$ fermionic creation and 
annihilation operators and $C$ is some constant. The dimensionless excitation 
energies $\Lambda_q^{}$ (which can be chosen to be positive and ordered, i.e., 
$\Lambda_N^{}\ge\ldots\ge\Lambda_2^{}\ge\Lambda_1^{}\ge0$, 
while properly adjusting the constant $C$)
satisfy the linear difference equations
\begin{eqnarray}
\Lambda_q \Psi_q(j) &=& -\, h_j^{} \Phi_q(j) - 
                        \varepsilon_j^{} \Phi_q(j+1) \; ,
\nonumber\\
\Lambda_q \Phi_q(j) &=& -\, \varepsilon_{j-1}^{} \Psi_q(j-1) - 
                        h_j^{} \Psi_q(j) \; .
\label{fermeq}
\end{eqnarray}
Normally, one proceeds by eliminating either $\Psi$ or $\Phi$.
In this way, the problem of diagonalising the 
$2^N\times 2^N$ matrix $H_{N}^{}$ can be reduced to the diagonalisation
of an $N\times N$ matrix. {}From the set of $N$ eigenvalues 
(fermion frequencies) $\Lambda_{q}^{}$ of the latter, the total 
spectrum of the complete 
Hamiltonian $H_{N}^{}$ is recovered by considering the sums of the
elements of any of its $2^N$ subsets. For periodic boundary conditions,
the situation is in fact a little more complicated; in order to avoid
the appearance of a non-local number operator in the fermionic Hamiltonian 
one has to consider so-called {\em mixed sector\/} Hamiltonians. 
These are obtained
by combining the {\em even\/} sector (with respect to the operator
$\prod_{j=1}^{N}\sigma_j^z$ that commutes with $H_N^{}$) of 
the Hamiltonian (\ref{eq:QC}) with {\em periodic\/} 
boundary conditions with the {\em odd\/} sector of the same Hamiltonian
with {\em antiperiodic\/} boundary conditions
(defined by $\sigma^{x}_{N+1}=-\sigma^{x}_{1}$), and vice versa 
\cite{BCS,G90}. 

The IQC is equivalent to the 2$d$ classical Ising model with a
layered modulation of the interactions \cite{Kogut}. 
The correspondence shows up through an extremely anisotropic limit 
(also called the $\tau$-continuum limit \cite{FS}) of the coupling constants 
of the latter, which is believed not to alter the critical behaviour. 
Explicitly, one establishes the following relations:
The {\em temperature}\/ $T$ of the classical model 
corresponds to the {\em transverse field}\/ $h$ of the 
quantum chain, the {\em free energy}\/ translates into the 
{\em ground-state energy}\/ and the counterpart of the 
{\em correlation length $\xi_{\|}$ parallel}\/ to the layers is the 
{\em inverse mass gap}\/ $(E_1 - E_0)^{-1}$ of the IQC. 
In our setting, the mass gap is just given by the modulus of the 
smallest fermionic excitation $\Lambda_1$ of (\ref{ffh}). 
According to finite-size scaling, we thus expect a scaling 
behaviour of $\Lambda_1$ at the critical point as 
\begin{equation}\label{eq:scal}
\Lambda_1(N) = m^{\nu_\|/\nu_\bot}_{} \Lambda_1(mN) =: m^z_{} 
\Lambda_1(mN) \, ,
\end{equation}
provided the correlation length scales with a power law at all. 
Hereby, the scaling exponent $z$ translates into the correlation 
length critical exponent $\nu_{\|}$
parallel to the layers in terminology of the 2$d$ statistical system, 
while $\nu_{\bot}$ retains its unperturbed Onsager value $\nu_{\bot}=1$.

The Hamiltonian (\ref{eq:QC}) is {\em critical\/} (in the sense that
the energy gap in the excitation spectrum vanishes, i.e.,
$\Lambda_1=0$) 
in the thermodynamic limit if the {\em geometric mean\/} 
(rather than the arithmetic mean) 
of the couplings, normalised by the variables of the transverse field,
is equal to one, i.e.,
\begin{equation}\label{eq:critcond}
    \lim_{N\rightarrow\infty}
    \left|\frac{\varepsilon_1^{}}{h_1}  \cdots 
     \frac{\varepsilon_N^{}}{h_N}\right|^{1/N} \, = \, 1 \, ,
\end{equation}
compare Ref.~\cite{P}.
The repetitive structure of the substitution sequences we are going to 
consider guarantees that the mean is well-defined \cite{MQ,LP}.
Let us now assume that the coupling constants $\varepsilon_{j}$
($j=1,2,\ldots ,N$) are chosen from a set of $n$ possible values
$\varepsilon_{m}^{}$ ($m=1,2,\ldots ,n$), the value $\varepsilon_{m}^{}$
occuring with a frequency $p_{m}^{}$ (where $\sum_{m=1}^{n} p_{m}^{} = 1$).
For simplicity, we take the field to be constant, i.e., $h_j^{}\equiv h=1$.
Then, condition (\ref{eq:critcond}) can be fulfilled by choosing 
$n$ positive real numbers
$r_1^{},r_2^{},\ldots, r_n^{}$ with 
$r_1^{}\cdot r_2^{}\cdot\ldots\cdot r_n^{}=1$,
and setting the couplings to
\begin{equation}\label{eq:para}
       \varepsilon_{m}^{} \, = \, r_m^{p/p_{m}^{}} \, ,
\end{equation}
where $p = p_{1}^{}\cdot p_{2}^{}\cdot \ldots \cdot p_{n}^{}$. 
Unless stated otherwise, this parametrisation of the critical surface 
will be used in the sequel.

\subsubsection*{Substitution sequences}

Let us now consider coupling constants $\varepsilon_{j}^{}$ drawn 
from a set of values $\varepsilon_{\alpha}^{}$, where the label 
$\alpha\in{\cal A}$ runs over the letters of a (finite) $n$-letter 
alphabet ${\cal A}$.
We choose the couplings $\varepsilon_{j}^{}$ 
according to sequences generated by iterated application
of a substitution rule $\varrho: \alpha \to w_\alpha$ on ${\cal A}$. 
For the introduction of the concept and some notation, consider a simple
example on a two-letter alphabet ${\cal A}=\{a,b\}$:
\begin{equation}
\varrho_{}^{(k)}:
\begin{array}{rcl}
a & \rightarrow & ab \\
b & \rightarrow & a^k
\end{array}\; ,\;\;\;\;\;\;\;
\bm{M}_{\varrho^{(k)}} = \left(\begin{array}{@{\,}r@{\;\;}r@{\,}}
1&k\\ 1&0\end{array}\right)\; ,\;\;\;\;\;\;\;
\lambda_{\pm}^{(k)}=\frac{1 \pm \sqrt{4k+1}}{2}\; .
\label{eq:sub}
\end{equation}
We start with an initial word, e.g. ${\cal W}_{0}^{}=a$, and define 
${\cal W}_{m+1}^{}=\varrho_{}^{(k)}({\cal W}_{m}^{})$.
$\bm{M}_{\varrho^{(k)}}$ denotes the corresponding substitution matrix 
(whose elements count the number of letters $a$ and $b$ in the words 
$w_a$ and $w_b$), which has the eigenvalues $\lambda_{\pm}^{(k)}$. 
The entries of the, statistically normalised, (right) eigenvector
${\bm v}_{\rm PF}^{}$, corresponding to the
Perron-Frobenius eigenvalue $\lambda_{\rm PF}^{} = \lambda_{+}^{(k)}$,
give the frequencies of the letters in the limit word ${\cal W}_{\infty}^{}$.
In order to ensure that, starting from any seed, the letter frequencies 
converge to unique, positive values in the limit chain, we restrict ourselves
to {\em primitive}\/ substitution rules $\varrho$ throughout this paper. 
For the corresponding substitution matrices $\bm{M}_\varrho$, this means 
that the entries of $\bm{M}_\varrho^\ell$ are {\em strictly positive}\/ 
for any integer $\ell>\ell_0$ (and a suitably chosen finite $\ell_0\ge 0$).
This is obviously true for the examples (\ref{eq:sub}).

Associated to a word ${\cal W}_{m}^{}$ of length (i.e.,
number of letters)  $N=|{\cal W}_{m}^{}|$, 
one obtains finite (closed) quantum chains with $N$ sites by setting 
$\varepsilon_{j}^{}=\varepsilon_{a}^{}$ whenever the 
$j$\/th letter of the word  ${\cal W}_{m}^{}$ is an $a$, and 
$\varepsilon_{j}^{}=\varepsilon_{b}^{}$ otherwise, the final letter
of ${\cal W}_{m}^{}$ determining the coupling between the last
and the first spin in the chain.
While $\lambda_{\rm PF}^{}$, the largest eigenvalue of the 
substitution matrix, describes the 
asymptotic scaling of the chain length with the
number of iterations $m$, $N\sim(\lambda_{\rm PF})^{m}$, 
the {\em fluctuation}\/ behaviour of the exchange 
couplings is connected to the second-largest eigenvalue $\lambda_2$ 
(where it is the {\em modulus}\/ of the eigenvalue that matters, hence
by ``second-largest eigenvalue'' we always refer to the eigenvalue with the
second-largest absolute value). In order to quantify the fluctuations, 
let us define
\begin{equation}
g_{\cal W}^{} = \sum_{\alpha \in {\cal W}} 
\left(\varepsilon_\alpha - \bar{\varepsilon}\right)
\end{equation}
where $\bar{\varepsilon}$ is the average coupling in the limit chain. 
Fluctuations stay {\em bounded} as long as $|\lambda_2|<1$, 
but for $|\lambda_2|>1$ they grow with iterated substitutions 
as a power of the chain length, $g_{\cal W}\sim N^\beta$, governed by the
so-called {\em wandering exponent}\/ $\beta$ that is given by
\begin{equation} \label{beta}
\beta = \frac{\log |\lambda_2|}{\log \lambda_{\rm PF}^{}} \, .
\end{equation}  
In the limiting case $|\lambda_2|=1$, $|g_{\cal W}^{}|$ 
stays constant when we prolong the chain by complete iteration
steps. Generically, however, finite fluctuations within the substitutes 
$w_\alpha$ of the letters $\alpha$ will increase the total fluctuation 
in the vicinity of these points by a certain amount $\delta$: 
$g_{\widetilde{\cal W}}^{} \ge g_{\cal W}^{} + \delta$ 
for some word $\widetilde{\cal W}$ with 
$\left| |\widetilde{\cal W}|-|\varrho({\cal W})|\right| \le
\Delta $ and fixed $\Delta$. By repeated use of this argument,
we thus expect fluctuations that {\em diverge logarithmically}\/ with 
the length of the chain (since the length grows exponentially with the 
number of iterations), see also Ref.~\cite{D90}. On the other hand, the same 
argument also shows that, in this case, the fluctuation behaviour is not
completely determined by the {\em substitution matrix}\/, but may also depend 
on the details of the actual {\em substitution rule}\/ \cite{GB94}; and 
for special choices the fluctuations may still stay bounded even though
the second-largest eigenvalue is $|\lambda_2|=1$.

Coming back to our examples (\ref{eq:sub}), 
let us concentrate on the three cases $k=1$ 
(which yields the famous {\em Fibonacci sequence}\/),
$k=2$ (the {\em period-doubling sequence}\/), and 
$k=3$ (the {\em binary non-Pisot sequence}\/).
As the second-largest eigenvalues of the substitution matrices 
$\bm{M}_{\varrho^{(k)}}$ fulfill 
$|\lambda_{-}^{(1)}|<1$,  $|\lambda_{-}^{(2)}|=1$, 
and $|\lambda_{-}^{(3)}|>1$, this includes an example of 
bounded fluctuations ($k=1$), a marginal case
with logarithmically divergent fluctuations ($k=2$) and an example
of strong fluctuations ($k=3$), hence all three 
interesting cases are illustrated \cite{GB94}. 
Another substitution chain that 
is referred to several times in the text is the celebrated
{\em Thue-Morse sequence}\/
\begin{equation}
\varrho_{}^{({\rm TM})}:
\begin{array}{rcl}
a & \rightarrow & ab \\
b & \rightarrow & ba
\end{array}\; ,\;\;\;\;\;\;\;
\bm{M}_{\varrho^{({\rm TM})}} = \left(\begin{array}{@{\,}r@{\;\;}r@{\,}}
1&1\\ 1&1\end{array}\right)\; ,\;\;\;\;\;\;\;
\lambda_{+}^{({\rm TM})}=2\; ,\;\;\;
\lambda_{-}^{({\rm TM})}=0\; .
\label{eq:tm}
\end{equation}

\section{Renormalisation transformation}
\setcounter{equation}{0}

Rather than reducing the set of linear equations (\ref{fermeq}) to 
$N$ dimensions, we proceed (as does Ref.~\cite{IT}) with a direct treatment
as a $2N$-dimensional eigenvalue problem of the real symmetric matrix
\begin{equation}  \label{fh}
{\cal H} = \left(
\begin{array}{@{}cccccccc@{}}
0&h_1^{}&0&0&0&\cdots&0&\varepsilon_N^{}\\
h_1&0&\varepsilon_1^{}&0&0&\cdots&0&0\\
0&\varepsilon_1^{}&0&h_2^{}&0&\cdots&0&0\\
0&0&h_2^{}&0&\varepsilon_2^{}&\cdots&0&0\\
\vdots&\vdots&\ddots&\ddots&\ddots&\ddots&&\vdots\\
0&0&\cdots&0&\makebox[0pt]{$h_{\scriptscriptstyle N-1}^{}$}&
0&\makebox[0pt]{$\varepsilon_{\scriptscriptstyle N-1}^{}$}&0\\
0&0&\cdots&0&0&\makebox[0pt]{$\varepsilon_{\scriptscriptstyle N-1}^{}$}&
0&h_N^{}\\
\varepsilon_N^{}&0&\cdots&0&0&0&h_N^{}&0
\end{array}\right)\; .
\end{equation}
Since ${\cal H}$ is symmetric and bipartite\footnote{It can be
represented in the block-form
${\cal H}=\left(\begin{array}{@{}c@{\;\;}c@{}}0&A\\A&0\end{array}\right)$ 
after an appropriate reordering of the basis.}, the eigenvalues 
occur in real pairs of opposite sign. But changing the sign of an 
eigenvalue simply means interchanging the corresponding
creation and annihilation operators in Eq.~(\ref{ffh}), 
thus we can restrict ourselves to the positive part of the spectrum.
In what follows, 
we concentrate on the case of a constant field $h_j^{}\equiv h$, 
where we may set $|h|=1$ by an appropriate choice of the energy scale. 
Of course, one cannot possibly expect analytical results for an arbitrary 
sequence of couplings. However, for {\em substitution}\/
sequences, more can be said using an exact renormalisation scheme, and,
for an interesting class of chains with marginal fluctuations, 
the scaling exponent $z$ (\ref{eq:scal}) can be determined exactly and 
given in a closed form.

Consider now sequences of couplings chosen according to
a substitution rule
\begin{equation} \label{subs}
\varrho :\;\; a_i \to a_i w_i 
\end{equation}
on an $n$-letter alphabet ${\cal A} = \{a_1, \ldots ,a_n\}$. Here,
the $w_i$ are $n$ arbitrary words of finite length in the alphabet ${\cal A}$.
Indeed, for the case of {\em aperiodic two-letter\/} substitution chains, 
this form imposes no restriction at all, since these can all be generated by 
substitution rules of the form (\ref{subs}). 
To see this, note that any two-letter
substitution rule $\bar{\varrho}$ can be transformed
appropriately by inner automorphisms, i.e., by replacing
$\bar{\varrho} \leadsto \varrho: a_i \to
\hat{w}^{-1}\bar{\varrho}(a_i)\hat{w}$ [where a natural choice of
the word $\hat{w}$ is the common beginning, if any, of the
words $\bar{\varrho}(a_1)$ and $\bar{\varrho}(a_2)$], and/or by considering 
$\bar{\varrho}^2$ instead. Obviously, this does not alter the limit chain.
Note that this is not possible for certain {\em periodic} chains, namely
those which are generated by a substitution rule that replaces all letters 
by the same word (the period), or by a multiple of it. 
It is no problem, however, to treat this case separately.

The eigenvalue problem for the matrix ${\cal H}$ 
(\ref{fh}) can be reformulated using transfer matrices in the usual way. 
Our renormalisation transformation (RT) is,
however, most conveniently described in terms of the star-product formalism, 
borrowed from scattering theory \cite{RED}. 
We therefore introduce new $S$-transfer matrices, called $S$-matrices 
from now on, by
\begin{equation}
\left(\begin{array}{@{\,}c@{\,}} \psi_{2j}\\ \psi_{2k+1}\end{array} \right) =
 {\cal S}_{j|k}
\left(\begin{array}{@{\,}c@{\,}} \psi_{2j+1}\\ \psi_{2k}\end{array} \right)
\end{equation}
with
\begin{equation}
{\cal S}_{j|k} = 
{\cal S}_{j|j+1} \ast {\cal S}_{j+1|j+2}
\ast \ldots \ast {\cal S}_{k-1|k} =
\starprod{\ell=j}{k-1} {\cal S}_{\ell|\ell+1} \; ,
\end{equation} 
where the star-product (\raisebox{-1.5pt}{\Large $\ast$}) of two 
$2\times2$ matrices is defined as
\begin{equation}\label{star}
\left(\begin{array}{@{\,}c@{\;\;}c@{\,}}
e&\bar{e}\\r&\rho
\end{array} \right) \ast
\left(\begin{array}{@{\,}c@{\;\;}c@{\,}}
\rho_1&r_1\\\bar{o}&o
\end{array} \right) = 
\left(\begin{array}{@{\,}c@{\;\;}c@{\,}} e&0\\0&o \end{array} \right)+
\frac{1}{1-\rho \rho_1}
\left(\begin{array}{@{\,}c@{\;\;}c@{\,}}
\bar{e} r \rho_1&\bar{e} r_1\\
\bar{o} r &\bar{o}r_1 \rho
\end{array} \right) \; .
\end{equation}
In our case, according to Eq.~(\ref{fh}), the elementary $S$-matrices are
\begin{equation}
\label{sj}
{\cal S}_{j|j+1} = \left(
\begin{array}{@{\,}cc@{\,}}
\varepsilon_{j}^{-1} \Lambda & -\varepsilon_{j}^{-1} h \\[1mm]
- \varepsilon_{j+1}^{-1} h & \varepsilon_{j+1}^{-1} \Lambda 
\end{array} \right) \; ,
\end{equation}
where $\Lambda$ denotes an eigenvalue\footnote{So far, we do not have to
impose any restriction here, thus $\Lambda$ may be {\em any}\/ eigenvalue
of ${\cal H}$. Only later, after expanding the renormalisation equations in 
powers of $\Lambda$ [Eq.~(\ref{hren}) and below], we shall restrict ourselves 
to eigenvalues that vanish in the limit of an infinite system size.} 
of ${\cal H}$. 

In order to establish the renormalisation scheme, we introduce a number of
additional parameters for these $S$-matrices. 
For each letter $a_i\in{\cal A}$, 
we define two asymmetry-coefficients $\kappa_{a_i}^\pm$ 
and a field variable $h_{a_i}$
that enter the $n^2$ different elementary $S$-matrices 
through\footnote{Note that this labeling of the fields differs 
from that used in Eqs.~(\ref{eq:QC}) and (\ref{fh}) in that the
field variables are now labeled in the same way as the coupling constants
(i.e., as the bonds) rather then by the sites of the chain.}
\begin{equation}
S_{a_i|a_j} = \left(
\begin{array}{@{\,}cc@{\,}} \varepsilon_{a_i}^{-1} \kappa_{a_i}^+ \Lambda
& -\varepsilon_{a_i}^{-1} h_{a_i}\\ -\varepsilon_{a_j}^{-1}h_{a_i}& 
\varepsilon_{a_j}^{-1}
\kappa_{a_i}^- \Lambda \end{array} \right) \; .
\end{equation}
The RT now just {\em reverses}\/ the substitution steps.
In our framework, this is simply done by building 
\raisebox{-1.5pt}{\Large $\ast$}-products of the $S$-transfer 
matrices that correspond to the words $w_i$ given by the substitution rule
(\ref{subs}). Schematically, this works in the following fashion
\[
\begin{array}{ccccc}
& \multicolumn{3}{l}{
\mbox{\footnotesize $S_{a_i^{}|w_i^1}$}\;
\mbox{\footnotesize $S_{w_i^1|w_i^2}$}
\hphantom{w\;\ldots\;w}
\mbox{\footnotesize $S_{w_i^{|w_i|}|a_j^{}}$}}
& \\
& \multicolumn{3}{l}{\hphantom{a}
\mbox{\LARGE $\curvearrowright$}\hphantom{w}
\mbox{\LARGE $\curvearrowright$}\hphantom{\;w\;\ldots\;w\;}
\mbox{\LARGE $\curvearrowright$}
} &  \\
\cdots&
\underbrace{\vphantom{w_j^1}\bm{a_i^{}}\;\;\;\; w_i^1\;\;\;\; w_i^2\; 
            \ldots\; w_i^{|w_i|}}_{}&
\underbrace{\vphantom{w_j^1}\bm{a_j^{}}\;\;\;\; w_j^1\;\;\;\; w_j^2\; 
            \ldots\; w_j^{|w_j|}}_{}& 
\underbrace{\vphantom{w_j^1}\bm{a_k^{}}\;\;\;\; w_k^1\;\;\;\; w_k^2\; 
            \ldots\; w_k^{|w_k|}}_{}& 
\cdots\\
&
\mbox{\large$\hphantom{\varrho}\,\uparrow\,\varrho$}& 
\mbox{\large$\hphantom{\varrho}\,\uparrow\,\varrho$}& 
\mbox{\large$\hphantom{\varrho}\,\uparrow\,\varrho$}& \\[1ex]
\cdots & \mbox{\Large \bm{a_i^{}}} & \mbox{\Large \bm{a_j^{}}} & 
         \mbox{\Large \bm{a_k^{}}} &\cdots \\
& \multicolumn{3}{c}{
\begin{picture}(50,30)(0,0)
\thicklines
\put(  0,20){\oval(100,35)[br]}
\put( 50,20){\vector(0,1){5}}
\end{picture}
\hphantom{\mbox{\Large $a$}}
\begin{picture}(116,30)(0,0)
\thicklines
\put(  0,20){\line(0,1){5}}
\put( 58,20){\oval(116,35)[b]}
\put(116,20){\vector(0,1){5}}
\put( 58,15){\makebox[0pt]{\Large $\tilde{S}_{a_i^{}|a_j^{}}$}}
\end{picture}
\hphantom{\mbox{\Large $a$}}
\begin{picture}(116,30)(0,0)
\thicklines
\put(  0,20){\line(0,1){5}}
\put( 58,20){\oval(116,35)[b]}
\put(116,20){\vector(0,1){5}}
\put( 58,15){\makebox[0pt]{\Large $\tilde{S}_{a_j^{}|a_k^{}}$}}
\end{picture}
\hphantom{\mbox{\Large $a$}}
\begin{picture}(50,30)(0,0)
\thicklines
\put(  0,20){\line(0,1){5}}
\put( 50,20){\oval(100,35)[bl]}
\end{picture}
} & 
\end{array}
\]
where $w_i^k$ denotes the $k$\/th letter of the word $w_i$. 
In this way, we obtain renormalised $S$-matrices
\begin{equation} \label{sren}
\tilde{\cal S}_{a_i|a_j} =
S_{a_i|w_i^1} \ast S_{w_i^1|w_i^2} \ast \cdots \ast S_{w_i^{|w_i|}|a_j}
=:\left(\begin{array}{@{\,}cc@{\,}}
\varepsilon_{a_i}^{-1} \tilde{\kappa}_{a_i}^+ \tilde{\Lambda} 
& - \varepsilon_{a_i}^{-1} \tilde{h}_{a_i}\\[1mm]
- \varepsilon_{a_j}^{-1} \tilde{h}_{a_i} 
& \varepsilon_{a_j}^{-1} \tilde{\kappa}_{a_i}^-\tilde{\Lambda} 
\end{array} \right) \; ,
\end{equation}
compare Eq.~(\ref{sj}).
As an {\em additional condition}\/ on the RT,
and in accordance with their initial values, we keep the product of the 
$2n$ asymmetry-coefficients $\kappa^{\pm}_{a_i}$ fixed:
\begin{equation}\label{kappacond}
\prod_{i=1}^n \tilde{\kappa}_{a_i}^+ \tilde{\kappa}_{a_i}^- = 
\prod_{i=1}^n \kappa_{a_i}^+ \kappa_{a_i}^- = 1 \; .
\end{equation}
Now, the RT for the different parameters can be read 
off directly from the \raisebox{-1.5pt}{\Large $\ast$}-product 
relations. Explicitly, we obtain the following recursion relations
\begin{eqnarray}
\label{RG1}
&
\makebox[ 5ex][r]{$\displaystyle h^{(1)}_{a_i}$} =
\makebox[10ex][l]{$\displaystyle h_{a_i}^{} \; ,$}\qquad
\makebox[ 5ex][r]{$\displaystyle h^{(k+1)}_{a_i}$} =
\makebox[25ex][l]{$\displaystyle 
                   \frac{-\varepsilon_{w_i^k} h_{w_i^k} h^{(k)}_{a_i}}
                        {\varepsilon_{w_i^k}^2 -\Lambda_{w_i^k+}^{} 
                         \Lambda_{a_i-}^{(k)}}\; ,$}\qquad
\makebox[ 5ex][r]{$\displaystyle\tilde{h}_{a_i}^{}$} =
\makebox[15ex][l]{$\displaystyle h^{(|w_i|+1)}_{a_i}\; ;$}
& 
\\[1mm]
\label{RG2}
&
\makebox[ 5ex][r]{$\displaystyle\Lambda_{a_i-}^{(1)}$} =
\makebox[10ex][l]{$\displaystyle\Lambda_{a_i-}^{}\; ,$}\qquad
\makebox[ 5ex][r]{$\displaystyle\Lambda_{a_i-}^{(k+1)}$} =
\makebox[25ex][l]{$\displaystyle\Lambda_{w_i^k-}^{} + 
                   \frac{h_{w_i^k}^2\Lambda_{a_i-}^{(k)}}
                        {\varepsilon_{w_i^k}^2 - \Lambda_{w_i^k+}^{} 
                         \Lambda_{a_i-}^{(k)}}\; ,$}\qquad
\makebox[ 5ex][r]{$\displaystyle\tilde{\Lambda}_{a_i-}^{}$} = 
\makebox[15ex][l]{$\displaystyle\Lambda_{a_i-}^{(|w_i|+1)}\; ;$}
&
\\[1mm] 
\label{RG3}
&
\makebox[ 5ex][r]{$\displaystyle\Lambda_{a_i+}^{(1)}$} =
\makebox[10ex][l]{$\displaystyle\Lambda_{a_i+}^{}\; ,$}\qquad
\makebox[ 5ex][r]{$\displaystyle\Lambda_{a_i+}^{(k+1)}$} =
\makebox[25ex][l]{$\displaystyle\Lambda_{a_i+}^{(k)} + 
                   \frac{\Lambda_{w_i^k+}^{}
                         \left[h^{(k)}_{a_i}\right]^2}
                        {\varepsilon_{w_i^k}^2 - \Lambda_{w_i^k+}^{}
                         \Lambda_{a_i-}^{(k)}}\; ,$}\qquad 
\makebox[ 5ex][r]{$\displaystyle\tilde{\Lambda}_{a_i+}^{}$} = 
\makebox[15ex][l]{$\displaystyle\Lambda_{a_i+}^{(|w_i|+1)}\; ;$}
&
\end{eqnarray}
for the field variables $h_{a_i}^{}$ and for $\Lambda_{a_i\pm}^{}$ 
that are defined as
\begin{equation}
\Lambda_{a_i\pm}^{} := \kappa_{a_i}^{\pm} \Lambda 
\; , \qquad
\tilde{\Lambda}_{a_i\pm}^{} := \tilde{\kappa}_{a_i}^{\pm} 
\tilde{\Lambda} \; .
\end{equation}
Using Eq.~(\ref{kappacond}), we obtain the renormalised fermion frequencies 
$\tilde{\Lambda}$ as 
\begin{equation} \label{kappacond2}
\tilde{\Lambda} =
\left(\prod_{i=1}^n 
\tilde{\Lambda}_{a_i+}^{}\tilde{\Lambda}_{a_i-}^{}\right)^{\frac{1}{2n}} \; .
\end{equation}
Note the following properties of our renormalisation approach:
\begin{itemize}
\item
The RT is {\em exact}, since no new parameters are introduced 
in a renormalisation step. Altogether, there are $3n$ parameters
for $n$ different couplings: Effectively $2n-1$ asymmetry-parameters 
$\kappa_{a_i}^{\pm}$, the $n$ fields $h_{a_i}$, and the eigenvalue $\Lambda$. 
Note that {\em the coupling constants are not renormalised}\/ in our scheme.  
\item
The rescaling factor of the system size is just given by the leading 
eigenvalue $\lambda_{\rm PF}^{}$ of the substitution matrix corresponding 
to the substitution rule under consideration.
\item
The critical point of the model, defined by the vanishing of the mass gap
($\Lambda \equiv 0$), corresponds to a fixed point of the RT 
for $\Lambda$, as can easily be seen from Eqs.~(\ref{RG2}) and (\ref{RG3}).
\item
The scaling behaviour in the vicinity of the critical point is governed by the 
leading-order term of the RT as $\Lambda\to 0$. 
\end{itemize}
We now proceed to establish the RT of the spectrum to 
leading order for $\Lambda \to 0$ explicitly. This is most easily done 
by noting that, in the product (\ref{sren}), the common prefactors 
$(1-\rho\rho_1)^{-1}=1+{\cal O}(\Lambda^2)$ 
of the \raisebox{-1.5pt}{\Large $\ast$}-products (\ref{star}) 
gives rise to higher-order corrections only; 
and therefore, for the present purpose, the denominator can be neglected 
completely. Consider first the renormalisation of the field variables. 
We obtain
\begin{equation} \label{hren}
\tilde{h}_{a_i} = h_{a_i} \cdot\prod_{\ell=1}^{|w_i|} \frac{-h_{w_i^\ell}}
{\varepsilon_{w_i^\ell}} + {\cal O}(\Lambda^2) \; .
\end{equation}
Instead of working directly with the RT for the 
fields, it is advantageous to consider new variables 
\begin{equation} \label{y0}
y_i := \log \left(\frac{h_{a_i}^2}{\varepsilon_{a_i}^2}\right)\; .
\end{equation}
To linear order in $\Lambda$, the RT of $y_i$ then yields the simple 
matrix form
\begin{equation} \label{y1}
\bm{\tilde{y}} =  \bm{M}_{\varrho}^t\, \bm{y} \; ,
\end{equation}
where $\bm{M}_{\varrho}^t$ is just the {\em transpose}\/ of the 
substitution matrix of $\varrho$. Now, take a look at the initial 
conditions ${\bm y}^{[0]}$ for ${\bm y}$ which are 
$y_i^{[0]} = - 2\log \varepsilon_{a_i}$. As stated above, 
the entries of the (statistically normalised) Perron-Frobenius eigenvector 
${\bm v}_{\rm PF}^{}$ are the asymptotic letter frequencies 
$p_i:=p_{a_i}$. We thus obtain 
\begin{equation} \label{y2}
{\bm y}^0 \cdot {\bm v}_{\rm PF}^{} =  - 2\log\left[\prod_{i=1}^{n}
\varepsilon_{a_i}^{p_i}\right] = 0
\end{equation}
for an {\em arbitrary}\/ set of couplings fulfilling the criticality
condition (\ref{eq:critcond}). Hence, ${\bm y}^{[0]}$ is nothing but an 
arbitrary (real) linear combination of the (right) 
eigenvectors of $\bm{M}_{\varrho}^t$ (hence of the {\em left}\/ eigenvectors
of $\bm{M}_{\varrho}$) {\em excluding}\/ the Perron-Frobenius one. Thus, 
for a generic choice of critical couplings, it scales with the second-largest
eigenvalue, $\|\bm{y}^{[k]}\| \sim |\lambda_2|^k$, under the iteration
${\bm y}^{[k+1]}=\bm{M}_{\varrho}^t {\bm y}^{[k]}$. 
Translated into the terminology of the renormalisation group, we recognise 
${\bm y}$ as a {\em scaling field}\/ transforming according to
\begin{equation}
\bm{\tilde{y}} = (\lambda_{\rm PF})^{\beta} {\bm y}
\end{equation} 
where the corresponding {\em renormalisation group eigenvalue}\/ is just 
the wandering exponent $\beta$ defined in Eq.~(\ref{beta}). This demonstrates
explicitly that power law, logarithmic, or bounded fluctuations lead to 
relevant, marginal, or irrelevant scaling fields, respectively. 
We are going to describe below how this affects the critical behaviour.

We now proceed to the RT for $\Lambda_{a_i\pm}$. Defining
\begin{eqnarray}
P_{a_i a_j}^+ &:=&
\sum_{k=1}^{|w_i|} \delta_{w_i^k,a_j}  \frac{h_{a_i}^2}{h_{w_i^k}^2} 
\prod_{\ell =1}^k 
\frac{h_{w_i^\ell}^2}{\varepsilon_{w_i^\ell}^2} + \delta_{a_i,a_j} \; ,
\\  \label{pminus}
P_{a_i a_j}^- &:=&
\left[\sum_{k=1}^{|w_i|} \delta_{w_i^{k},a_j} \prod_{\ell = 1}^{k} 
\frac{\varepsilon_{w_i^\ell}^2}{h_{w_i^\ell}^2} + \delta_{a_i,a_j}\right]
\prod_{\ell =1}^{|w_i|} \frac{h_{w_i^\ell}^2}{\varepsilon_{w_i^\ell}^2} \; ,
\end{eqnarray}
we find, by induction,  as the linear part of the RT of $\Lambda_{a_i\pm}$
\begin{equation}
\tilde{\Lambda}_{a_i\pm} = \sum_{j=1}^n \Lambda_{a_j\pm} P_{a_i a_j}^\pm\; .
\end{equation}
Setting $M^\pm_{ij}:=P^\pm_{a_ia_j}$, this can also be written in a matrix
form
\begin{equation}
\bm{\tilde{\Lambda}}_\pm = \bm{M}^\pm \bm{\Lambda}_\pm\; .
\end{equation}
Since all the components of the matrices $\bm{M}^\pm$ and of the vectors
$\bm{\Lambda}_\pm$ are positive, the vectors converge to the 
Perron-Frobenius eigenvectors of $\bm{M}^\pm$ under iteration of the RT. 
Let $\mu^\pm$ denote the Perron-Frobenius eigenvalues of $\bm{M}^\pm$. 
{}From Eq.~(\ref{kappacond2}), we conclude that
\begin{equation}
\tilde{\Lambda}^2 = \mu^+ \mu^- \Lambda^2 \; .
\end{equation}
For further convenience, we transform $\bm{M}^+$ (under conservation 
of the spectrum) according to $\bm{M}^{+}_T = \bm{T}^{-1} \bm{M}^+ \bm{T}$
with a diagonal transformation matrix $T_{ij} = h_{a_i}^2\delta_{ij}$. 
As new entries, we find, dropping the index $T$, 
\begin{equation} \label{pplus}
M_{ij}^{+} = 
P_{a_ia_j}^+ =
\sum_{k=1}^{|w_i|} \delta_{w_i^k,a_j}  \prod_{\ell =1}^k 
\frac{h_{w_i^\ell}^2}{\varepsilon_{w_i^\ell}^2} + \delta_{a_i,a_j} \; .
\end{equation}
This form shows explicitly that, to linear order in $\Lambda$,
only the {\em reduced couplings}\/ $\varepsilon_{a_i}/h_{a_i}$ enter in
the RT. Finally, since the rescaling factor of the chain length is the
largest eigenvalue $\lambda_{\rm PF}^{}$ of the 
substitution matrix, we formally obtain the scaling exponent $z$ through
\begin{equation} \label{zz}
\Lambda_j = x_j N^{-z} \; , \qquad z =  \frac{\log \left(\mu^+
\mu^-\right)}{2 \, \log \left(\lambda_{\rm PF}^{}\right)}\; .
\end{equation}
Let us now look at the different cases of fluctuation behaviour in more detail.

\subsection{Bounded Fluctuations}
Because $\bm{y}$ is perpendicular to the Perron-Frobenius eigenvector
of the substitution matrix, and since all other eigenvalues are smaller than
one in modulus, we find $\|\bm{y}^{[k]}\|\to 0$ for all substitution
chains with bounded fluctuations, hence $h_{a_i}^2/\varepsilon_{a_i}^2 \to 1$.
Evaluating the transformation matrices $\bm{M}^\pm$ at this fixed point of
the fields, we obtain from Eqs.~(\ref{pminus}) and (\ref{pplus})
\begin{equation}
 \bm{M}^+ =  \bm{M}^- = \bm{M}_{\varrho}
\end{equation}
and hence
\begin{equation}
\mu^+ = \mu^- = \lambda_{\rm PF}^{} \; .
\end{equation}
Thus we have $z = 1$ for substitution chains with bounded fluctuations, and 
the low-energy spectrum at criticality scales as
\begin{equation}
\Lambda_j = x_j  \frac{v}{N} \, .
\end{equation} 
Hereby, the $\Lambda$-transformation induces a finite renormalisation of the
{\em fermion velocity}\/ $v$ in comparison to the uniform chain
\begin{equation} \label{vren}
v = \lim_{k \to \infty \atop \Lambda \to 0 } 
\frac{(\lambda_{\rm PF})^k}{\Lambda^{[k]}/\Lambda}
\end{equation}
where $\Lambda^{[k]}$ is the image of $\Lambda^{[0]}=\Lambda$ after a
$k$-fold application of the RT. 

In general, all renormalisation steps contribute in Eq.~(\ref{vren}). 
However, if {\em all}\/ eigenvalues of the substitution matrix but the 
Perron-Frobenius one are zero, the fixed point of the RT is already 
reached after the first renormalisation step and $v$ is obtained 
immediately. This is shown below for periodic chains, but also allows to
calculate the fermion velocity of the Thue-Morse and related chains. 
Using the equidistribution property of quasiperiodic {\em cut-and-project}\/
chains, like our first example (\ref{eq:sub}) with $k=1$ (the Fibonacci
sequence), the limit 
in Eq.~(\ref{vren}) can be carried through explicitly for these cases, too. 
Since this has already been done in Ref.~\cite{L}, we do not repeat the 
arguments here, and merely cite the result
\begin{equation}
v(r) =\frac{2\log(r)}{r-r^{-1}} 
\end{equation}
for two couplings parametrised as in Eq.~(\ref{eq:para}).

\subsubsection*{Periodic chains}

In this short part, we briefly show how the known results for the 
fermion velocity $v$ for {\em periodic}\/ quantum chains can be recovered 
easily within our renormalisation scheme. By a substitution rule of the form
\begin{equation}
\varrho:\;\; a_i \to w \; ,
\end{equation}
{\em any}\/ periodic chain with period $m=|w|$ can be generated.
In this simple situation, we can take all the fields $h_{a_i}\equiv h$ 
and asymmetry-parameters 
$\kappa_{a_i}^+\equiv\kappa\equiv(\kappa_{a_i}^-)^{-1}$ to be equal. 
By virtue of the criticality condition (\ref{eq:critcond}), 
the reduced couplings transform 
as $h^2/\varepsilon_i^2 \to 1$ [see Eq.~(\ref{hren})],  
and we obtain the (fermionic) Hamiltonian of a {\em uniform} chain 
($\varepsilon = h = \varepsilon_{w^1}$) after a single 
renormalisation step. Note that the additional asymmetry-parameters 
result in an irrelevant similarity transformation of
the matrix (\ref{fh}) only. The renormalisation factor for 
$\Lambda$ can be deduced from Eqs.~(\ref{pminus}) and (\ref{pplus}), 
explicitly it reads (using Eq.~(\ref{eq:critcond}), and normalising to 
$\varepsilon=h=1$):
\begin{eqnarray} \nonumber
m^2 v^{-2} := \frac{\tilde{\Lambda}^2}{\varepsilon_1^2\Lambda^2} &=& 
\left( 1 + \sum_{k = 2}^{|w|} 
\varepsilon_2^2 \cdots \varepsilon_k^2 \right) \left(1 + \sum_{k = 2}^{|w|} 
\varepsilon_2^{-2} \cdots \varepsilon_k^{-2} \right) \\[2mm] \label{v}
&=& \sum_{k=1}^{|w|}\sum_{l=1}^{|w|} \varepsilon_k^2 \cdots 
\varepsilon_{k+l-1({\rm mod}\,|w|)}^2 
\end{eqnarray}
with $\varepsilon_{\ell}^{} := \varepsilon_{w^{\ell}}^{}$. 
The critical scaling of the fermion frequencies now results in 
\begin{equation}
\Lambda_j = \frac{2\pi j}{N}\,v \;,\qquad j \ll N\; ,
\end{equation}
where $v$ is given in Eq.~(\ref{v}), and where the well-known 
scaling behaviour of the uniform chain is recovered for $v=1$. 
This result was first derived in the context of 
the 2$d$ classical Ising model with layered periodicity in Ref.~\cite{AM}.

In general, there is no closed expression for the fermion velocity $v$. 
For the special case of a uniform chain of $\varepsilon_a$-couplings 
with periodic $\varepsilon_b$-defects on every $m$\/th site, we derive
\begin{equation} \label{vper}
v(r) = \frac{\sinh[\log (r)/m]}{\sinh [\log (r)]/m} \; .
\end{equation}
Note that $v(r)=1$ for $m=1$ (simple periodic chain), but also
$v(r) \rightarrow 1$ for $m\rightarrow\infty$ since the parametrisation
(\ref{eq:para}) implies that $r\rightarrow 1$ in this limit.

\subsection{Marginal fluctuations}

We now consider the case where the second-largest eigenvalue(s) 
of the substitution matrix (and of its transpose) have $|\lambda|=1$.
Let $E_1$ denote the corresponding $m$-dimensional (joined) eigenspace of 
$\bm{M}_\varrho^t$, where $1\le m \le n-1$.
By Eqs.~(\ref{y1}) and (\ref{y2}), 
$\bm{y}^{[k]}$ converges to the projection of $\bm{y}^{[0]}$ on $E_1$ under
iteration of the RT. 
The critical surface being $(m-1)$-dimensional, 
we can parametrise $\bm{y}=\bm{y}^{[\infty]}$ in the renormalisation limit as
\begin{equation} \label{yform}
\bm{y}=\sum_{i=1}^m \log (r_i) \, \bm{v}^i
\end{equation}
with $m$ vectors $\bm{v}^i$ that span $E_1$ and coefficients $r_i >0$,
$i=1,2,\ldots,m$. This implies
\begin{equation}
\frac{h_{a_j}^2}{\varepsilon_{a_j}^2}  = \exp (y_j)  = \prod_{i=1}^m
r_i^{v_j^i}
\end{equation}
and, if the vector components $v^i_j$ are chosen as integers, 
the entries of the 
matrices $\bm{M}^\pm$ are polynomials with positive coefficients
of $m$ parameters $r_1^{\pm 1}, \ldots, r_m^{\pm 1}$.
If all $r_i$ are equal to 1, we obviously obtain
\begin{equation}
\bm{M}^+(r_i\!=\!1) = \bm{M}^-(r_i\!=\!1) = \bm{M}_\varrho  \; .
\end{equation}
Furthermore, if $\bm{y}^{[k]}$ converges to an eigenvector of 
$\bm{M}_\varrho^t$ to the eigenvalue $\lambda_2=1$,
$\bm{M}^+$ and $\bm{M}^-$ are related by
\begin{equation}
M^+_{ij}(r_1,\ldots,r_m) = M^-_{ij}(r_1^{-1},\ldots,r_m^{-1}) \;.
\end{equation}
To obtain the scaling exponent $z$ from Eq.~(\ref{zz}), we need to know the 
Perron-Frobenius eigenvalues $\mu^\pm$ of $\bm{M}^\pm$.
There is no explicit solution to that problem in the general case. 
In what follows, we present the complete answers for a 
number of more special, yet infinite, classes of substitution rules.

\subsubsection*{Cyclic permutations}

Let $w$ be any word from the alphabet, and $p_k(w)$ its $k$\/th cyclic 
permutation. Consider now a substitution rule $\varrho$ of the form
(\ref{subs}) with
\begin{equation} \label{rhoform2}
w_i = \left[ p_{k_i}(w) \right]^{l_i} \; .
\end{equation}
One special case is to take equal words $w_i=w$ for all $i=1,2,\ldots,n$. 
The corresponding substitution matrices indeed have $\lambda_2 = 1$ 
as their second-largest eigenvalue, with degeneracy $(n-1)$, 
since the column vectors of $\bm{M}_\varrho-\mathbb{I}$ differ 
by scalar factors only. Thus the fields stay at their original values 
and are not renormalised at all. To derive $\mu^\pm$ as a function of the 
parameters $r_1, \ldots, r_{n-1}$, we simply note that the special form of 
$\varrho$ transforms this property to $\bm{M}^\pm$, not just for the point 
$r_i =1 ,\,\forall i$, 
but {\em independently}\/ of the parameters. We thus obtain
\begin{equation}\label{mu}
\mu^\pm = \mbox{tr}(\bm{M}^\pm) - n +1 > 1 
\end{equation}
and the scaling exponent is given by
\begin{equation}
z = \frac{\log\left(1-n+\sum_{i=1}^n P_{a_ia_i}^+\right)+
\log\left(1-n+\sum_{i=1}^n P_{a_ia_i}^-\right)}{2\log 
(\lambda_{\rm PF}^{})} \; ,
\end{equation}
where the $P^\pm_{a_i,a_j}$ are given in Eqs.~(\ref{pminus}) 
and (\ref{pplus}) as functions of the couplings.
Since $\mu^\pm$ are polynomials of the parameters $r_i^{\pm1}$ with only 
positive coefficients, and $\mu^+(\{r_i\}) = \mu^-(\{r_i^{-1}\})$, we conclude 
$z \ge 1$. As a function of the parameters, $z$ is either purely 
{\em convex}\/, with unique minimum $z(r_i=1)=1$ (the limit of the 
uniform chain), or it is {\em constant}\/: $z \equiv 1$. It is interesting 
to take a closer look at the latter case. Obviously, this means that 
$P_{a_ia_i}^\pm = [\bm{M}_\varrho]_{ii}$, independently of the parameters. 
As can be shown, for substitution rules of the form (\ref{subs}) with 
(\ref{rhoform2}), 
this is possible if and only if $w$ is a word (or a power of a word)
with all letters $a_i$ appearing precisely once and if the 
last letter of $w_i$ is $a_i$. This means, however, that the resulting 
chain is {\em periodic}\/ and has bounded rather than logarithmically 
diverging fluctuations. 

\subsubsection*{Two-letter substitution rules}

As stated above, each two-letter substitution chain can be generated by
a substitution rule of the form (\ref{subs}). For any marginal substitution
rule, we can assume that the second-largest eigenvalue of the substitution 
matrix is $\lambda_2 = +1$. If necessary, this can be achieved by going
to $\varrho^2$. As a consequence, $\bm{y}^{[0]}$ in Eq.~(\ref{y1}) 
is eigenvector to the eigenvalue $\lambda_2=1$ of $\bm{M}_\varrho^t$, and 
as above the fields are already at their fixed points.

In this case, the entries of the transformation matrices $\bm{M}^\pm$ are 
polynomials of a single parameter $r$ that determines the critical couplings.
The crucial point now is to show that -- like in the case discussed 
above -- the matrices $\bm{M}^\pm$ have an eigenvalue 
$\lambda_2 =1$ independently of 
$r$ and of the detailed form of the substitution rule. 
This assertion is proved in Appendix~A. Then,  
the Perron-Frobenius eigenvalues of $\bm{M}^\pm$ are given 
as above (\ref{mu}), and we obtain the critical scaling exponent
\begin{equation} \label{z}
z(r) =  \frac{\log(P^+_{aa}+P^+_{bb}-1)+\log(P^-_{aa}+P^-_{bb}-1)}{2\log 
(\lambda_{\rm PF}^{})}
\end{equation}
where $\lambda_{\rm PF}^{}$ is the leading eigenvalue of the 
substitution matrix. As in the case 
discussed above, $z(r)$ is constant ($z\equiv 1$) only for the special case 
of a periodic chain (with period $ab$). Otherwise, $z$ is a convex 
function of $r$, with a unique minimum $z(1)=1$ where the (reduced) 
couplings are equal. On the other hand, it is easy
to see that the periodic chain with period $ab$ is the only 
two-letter substitution chain with bounded letter fluctuations
that can be generated by a substitution rule of the form (\ref{subs}) and
that has $\lambda_2 =1$ as the second-largest eigenvalue of the 
substitution matrix.  We thus conclude that $z\equiv 1$ {\em if and only
if}\/ the fluctuations are bounded, and $z>1$ otherwise.

\subsubsection*{Examples}

The {\em period-doubling chain}\/, given by the substitution rule
(\ref{eq:sub}) with $k=2$, falls into both classes described above.
We consider double substitution steps, eliminating blocks of the form $aba$ 
and obtain the scaling exponent
\begin{equation}
z(r) = \frac{\log(r^{1/3}+r^{-1/3})}{\log (2)}
\end{equation}
This result was reported before in Ref.~\cite{IT}.
If the length of the chain is increased by simple substitution steps, 
the coefficients $x_j$ in Eq.~(\ref{zz}) converge {\em separately}\/ 
for an even and an odd number of steps. Effectively, this can be seen 
as a renormalisation of the couplings according to 
$r = \varepsilon_a/\varepsilon_b \to r^{-1}$ in a single renormalisation 
step. Hence we obtain $x_j^{\rm odd}(r) = x_j^{\rm even}(r^{-1})$.

As an example that is not contained in the above classes, consider 
the three-letter substitution
\begin{equation}
\varrho:\; \begin{array}{ccl} a &\to& abc\\ b&\to& ba\\ c &\to&
ca\end{array}
\end{equation}
with eigenvalues of the substitution matrix $\lambda_{1,3}=1\pm\sqrt{2}$ 
and $\lambda_2 = 1$. The eigenvector to eigenvalue 1
of the transposed substitution matrix $\bm{M}_\varrho^t$ is $(0,t,-t)$,
and with Eq.~(\ref{y0}) we conclude that $h_a^2 \to \varepsilon_a^2$ and
$h_{b,c}^2 \to \varepsilon_b\varepsilon_c$.
Evaluating the RT of $\bm{\Lambda}_\pm$ at this fixed point, 
we obtain the critical scaling exponent 
\begin{equation}
z =  \frac{\log \left(1+\sqrt{1+\varepsilon_c/\varepsilon_b}\right) +
\log \left(1+\sqrt{1+\varepsilon_b/\varepsilon_c}\right)}{2\,\log
(1+\sqrt{2})}
\end{equation}
Note that, for $\varepsilon_b=\varepsilon_c$, the fluctuations of the
couplings are indeed bounded, and we obtain $z=1$ as it should be.
In our last example, the couplings follow the Thue-Morse sequence
on the {\em sites} of the chain, rather than on the bonds as in all 
cases discussed so far. 
The coupling constants for a site-problem are given as a function of 
the two adjacent bonds -- and thus can take in 
general $n^2$ different values ($2^2 =4$ in this case). It can, however, 
easily be shown that this can be reformulated as a bond-problem 
with a substitution chain on an $n^2$-letter alphabet \cite{MQ,BGJ,TBB}. 
In this case, the resulting four-letter substitution rule is 
\begin{equation}
\varrho:\; \begin{array}{ccc} a&\to& acdb\\ b&\to &bcdb\\ 
c& \to &cbac\\ d &\to& dbac \end{array} \; .
\end{equation}
The asymptotic letter frequencies are $p_a = p_d = 1/6$ and $p_b = p_c = 1/3$; 
and so we have $\varepsilon_a \varepsilon_d\varepsilon_b^2\varepsilon_c^2 =1$ 
(for $h = 1$) as the criticality condition (\ref{eq:critcond}). 
A possible parametrisation is 
$\varepsilon_a = q^2$, $\varepsilon_d = r^2$,
$\varepsilon_b = s$, and $\varepsilon_c = (qrs)^{-1}$.
The eigenvalues of the substitution matrix are $4$, $1$, $1$, and $0$, 
we thus expect marginal fluctuations, whereas the Thue-Morse bond-problem 
has bounded fluctuations.
Following the above steps, we obtain a scaling exponent
\begin{equation}
z = \frac{\log \left[ (qr)^{1/4} + (qr)^{-1/4}\right]}{\log (2)} \; .
\end{equation}
Note that for $\varepsilon_a=\varepsilon_d$, $\varepsilon_b=\varepsilon_c$ 
we obtain another coding of the period-doubling chain, while the special case
$\varepsilon_a=\varepsilon_b$, $\varepsilon_c=\varepsilon_d$ leads us back 
to the Thue-Morse bond-problem with bounded fluctuations 
(with $q=r^{-1}$, and thus $z=1$).

\subsection{Relevant fluctuations}

For $|\lambda_2|>1$, fluctuations diverge with a power law. As a 
consequence, the vector $\bm{y}^{[k]}$ finally scales with $|\lambda_2|$ 
under the RT. Since the matrix elements of $\bm{M}^\pm$ are proportional to
$\exp(y_i)$, at least a part of them eventually diverge like 
$\exp(c|\lambda_2|)$ [note that the components of $\bm{y}$ have different 
sign because of Eq.~(\ref{y2})].
Estimating the product of the Perron-Frobenius eigenvalues 
$\mu^\pm$ by the product of the corresponding traces, we obtain
\begin{equation}
\tilde{\Lambda}^2 \sim \tilde{\mu}^+ \tilde{\mu}^- 
\sim \Lambda^{2 |\lambda_2|}
\end{equation}
and thus
\begin{equation}
\Lambda \sim \exp\left( - c N^{\beta}\right)
\end{equation}
where the {\em wandering exponent}\/ $\beta$ is defined in Eq.~(\ref{beta}).

\subsubsection*{Two-letter substitutions}
Let us discuss this in more detail for the case of two-letter substitutions. 
We take $\lambda_2>1$ with corresponding eigenvector $\bm{v}$ of 
$\bm{M}_\varrho^t$ and parametrise the critical couplings as
\begin{equation}
\frac{h_{a,b}^2}{\varepsilon_{a,b}^2} = r^{v_{a,b}} \; .
\end{equation}
Since $v_{a}$ and $v_{b}$ have different sign, 
either one of these two is bigger than one for a fixed value $r \neq 1$, 
and without loss of generality we assume $r^{v_a}>1$.
Note that we have the following relation
\begin{equation} \label{product}
\prod_{\ell =1}^{|w_a|} \frac{h^2_{w_a^\ell}}{\varepsilon^2_{w_a^\ell}} = 
r^{(\lambda_2-1)v_a} \;.
\end{equation}
We now show that $P^\pm_{aa} \ge \exp\left[(\lambda_2-1)v_a\log r\right]$.
{}From Eq.~(\ref{pminus}), this is immediately clear for $P^-_{aa}$, but 
it also holds for $P^+_{aa}$. To see this, consider the products of 
subsequent couplings of the word $w_a$. The product of the entire word 
is given by Eq.~(\ref{product}). If the last letter of $w_a$ is an $a$, 
this enters $P^+_{aa}$ and we are done. On the other hand, if the last 
letter is a $b$, consider the product of all couplings but the last 
$h_b^2/\varepsilon_b^2$, which is even larger. Iterating this argument, 
we finally arrive at a letter $a$ which proves our estimate of $P^+_{aa}$. 
Using the standard eigenvalue formula,
we obtain the scaling of $\mu^\pm$ under the RT, and thereby
\begin{equation}
\Lambda \sim \exp\left( -c |\log r| \, N^{\beta}\right) \;.
\end{equation}
Here, $c$ is some constant that may depends on the sign of $\log(r)$, 
but apart from that is independent of $r$ and $N$. This scaling form 
was also predicted in \cite{L}.

\section{Scaling behaviour and critical properties}
\setcounter{equation}{0}

In this section, we give a short summary of the results of our renormalisation
treatment and discuss their consequences for the critical properties of the
IQC.

\subsection{Results: Bounded fluctuations}

Here, the behaviour of the low-energy part of the the spectrum at
criticality is essentially the same as for the uniform chain. 
The correlation length critical exponents stays at its Onsager 
value $z = \nu_{\|} = \nu_{\bot} =1$. For
periodic chains (or periodic variations of the interactions along the
layers of the 2$d$ statistical system), the results from renormalisation show that the
smallest fermion frequencies are only altered by a common factor $v$ 
(fermion velocity), thus also conformal properties persist and the
whole model is described by the central charge $c=1/2$ conformal field
theory of a massless free Majorana fermion:
\begin{equation} \label{lam1}
\Lambda^{}_{j}\; \sim\; x_{j}^{} \; \frac{\pi v}{N}
\end{equation}
with a fermion velocity $v$ given by (\ref{v}).The
scaling dimensions $x_{j}^{}$ are odd integers
for periodic boundary conditions (mixed sector Hamiltonian) 
and odd half-integers for free 
boundaries \cite{C}.
Numerical results \cite{GB94} confirm this conformal behaviour 
even for non-periodic chains with bounded fluctuations.

\subsection{Results: Logarithmically diverging fluctuations}

For the critical scaling, we can make the following {\it Ansatz\/}:
\begin{equation}
\Lambda^{}_j \ = \left(\frac{j}{N}\right)^{z(r)} F_r(j,N)
\end{equation}
The index $r$ here denotes the dependence on the position on the
critical surface. The predicted scaling behaviour \cite{L} is confirmed 
analytically for all marginal two-letter substitution chains and some 
infinite classes of $n$-letter substitution chains. 
$j^z F_r(j,N)$ converges to the scaling coefficients $x_j$
if we increase the length of the chain $N$ by (multiple) substitution
steps, in other words $F_r(j,\log N)$ is asymptotically periodic for $N
\gg 0$ with period $\log \lambda_{\rm PF}^{}$, 
where $\lambda_{\rm PF}^{}$ is the scaling factor of the
system size. Eq.~(\ref{z}) implies $1 \le z < \infty$, and $z=1$ is only
obtained in the limit of the uniform chain where the fluctuations
actually vanish.
We further note that, in the explicitly solved cases, $z$ is invariant under
inversion of all couplings. Thus this is a symmetry of the lower part
of the critical spectrum, while it is easy to see that it is not a
symmetry of the entire finite-size spectrum. As a consequence, $z$
quadratically approaches the isotropic value $z_0 = 1$ for small
deviations of the couplings from the uniform strength
$\varepsilon_\alpha = 1$, as was also predicted by Luck~\cite{L}.

In the terminology of the 2$d$ classical Ising model, 
we have $\nu_\|=z >1$ for any marginal
disorder within layers of couplings. 
Since the correlation length critical exponent
perpendicular to the layers remains $\nu_\bot =1$, 
marginal fluctuations introduce a
relevant anisotropy of the model, as was noticed in Ref.~\cite{BBHILMT}. 
Due to anisotropic hyperscaling, a {\em negative}\/ specific heat exponent
\begin{equation}
\alpha = 2 - \nu_\| - \nu_\bot = 1 - z
\end{equation}
is expected.

Within the renormalisation approach, the length of the chain $N$ is
increased by (multiple) substitution steps. Thus, it is a priori not
clear what happens to the scaling dimensions when
we increase $N$ arbitrarily. Indeed,
numerical studies (for free boundary conditions) show that, 
if we increase $N$ by arbitrary amounts, while $z$ converges, 
$F_r(j,N)$ stays finite but will not tend to
a unique limit. This is in contrast to the bounded fluctuation case.
In more detail, the following properties of $F_r(j,N)$ have been
observed in all examples (of two-letter substitution chains) studied:
\begin{itemize}
\item Given the couplings, $F_r(j,N)$ appears to be bounded for all $N$ and
$j$.
\item For $N > j \gg 0$, the effect of $j$ on $F_r$ is a mere shift:
$F_r(j,N) \sim F(\log (N/j))$.
\item As a consequence of the RT, $N \gg j$, $F_r(j,\log N)$ is 
asymptotically periodic. 
\item The amplitude of $F_r$ vanishes in the limit of the
uniform chain, $r \to 1$, as $F_r$ approaches continuously the 
constant value $\pi$.
\end{itemize}
These points can immediately be translated into properties of the
integrated density of states (IDOS), the inverse of which is given by
\begin{equation}
\sigma^{-1}(\rho) = \rho^z F_r(-\log \rho)\;,\quad 0 < \rho \le 1 .
\end{equation}
While the first two points imply the convergence of this quantity,
the third point above implies the lower part of the IDOS to be 
self-similar with a discrete scaling symmetry
\begin{equation}
m \sigma(\omega) = \sigma(m^z\omega) = \sigma(\omega/v)
\end{equation} 
where $\log m$ is the asymptotic period of $F_r$. Mind that in general not the
spectrum (the lowest fermion frequencies), but only the IDOS displays 
this scaling behaviour for $N \gg 0$.

This scaling property has consequences for the possible gap
structure of the critical spectrum, where the results of the renormalisation
approach at least lead to necessary conditions for gap-labelling theorems. 
But since within this class more spectral properties can
be found analytically using trace maps, we postpone the discussion to
a forthcoming publication.  

\subsection{Results: Relevant fluctuations}

In substitution sequences with strong fluctuations, as in our third
example, the binary non-Pisot sequence [$k=3$ in Eq.~(\ref{eq:sub})], 
their divergence is described \cite{L} by the wandering exponent 
$\beta$ (\ref{beta}) which determines the exponential closing of the 
spectral gaps as
\begin{equation}
\Lambda^{}_{j} \;\sim\;
\left( A(r)\, +\, B(r)\, N^{-\alpha_{j}^{}(r)}\right)\;
\exp\left(-F_r(j,N)
\Delta(r)\, \left(\frac{N}{j}\right)^{\beta}\right) \; .
\end{equation}
Here, $\Delta(r)$ is the second moment of the distribution 
of coupling constants along the chain \cite{L} which for the
binary non-Pisot sequence is given by
\begin{equation}
\Delta(r)\; =\; \sqrt{2\sqrt{13}-7}\;\;\:\left|\log{(r)}\right|
\; \approx \; 0.459 \:\left|\log{(r)}\right| 
\end{equation}
and compatibility with the uniform case requires
$A(1)=0$ and $\alpha_{j}^{}(1)=1$. 

{}From the  results of the renormalisation approach, 
$F_r(j,\log N)$ should be asymptotically periodic for $N \gg j$.
This is confirmed by numerical investigations. 
Indeed, the latter suggest the very same properties 
as given above in the case of marginal fluctuations. In particular, we
observe $F_r(j,N) = F_r(\log (N/j))$ for $j \gg 0$.

Again, this means that the integrated spectral density converges in
the thermodynamic limit and its lower part shows the following scaling
property at the critical point:
\begin{equation}
\sigma (\omega^{(m^{-\beta})}) = m \sigma (\omega)\;,\quad \omega \ll
1 \;.
\end{equation}

\section{Concluding remarks}
\setcounter{equation}{0}

Using a renormalisation approach, we obtained
analytic results for the critical behaviour 
of a class of Ising models intermediate between uniformly 
(or periodically) ordered and randomly disordered systems. 
The systems under consideration are ferromagnetic Ising quantum chains whose 
interaction constants are modulated according to substitution sequences. 
Luck's criterion \cite{L} for the relevance of aperiodic disorder on the 
critical behaviour is fully confirmed for these systems, 
bringing along increasing evidence that the 
{\em fluctuation of the interactions}\/ is the basic concept that demarcates
the Onsager universality class from models with weaker critical singularities.
For two-letter substitution rules, this analysis was carried out in detail,
showing how the renormalisation flow is determined by the nature 
of the fluctuations. An exact renormalisation formalism for the most 
general case of $n$-letter substitution rules was also presented and 
applied to examples. 

A number of quantitative results have been displayed, mainly for the 
correlation length critical exponent (resp.~the scaling exponent $z$ 
of the mass gap), confirming and refining Luck's predictions for bounded, 
marginal, and relevant fluctuations of the coupling constants. In particular, 
an exact formula has been obtained for the scaling exponent of arbitrary 
two-letter substitution chains, containing previous results as special cases.
In this light, the case of two-letter substitution rules appears to be
rather well understood by now.

What remains to be done in the $n$-letter case is a thorough discrimination 
between quantum (substitution) chains with bounded and 
with marginal fluctuations of the couplings. 
This is connected with the problem of determining all substitution rules with
second-largest eigenvalue of the substitution matrix on the unit circle, 
which nevertheless result in a chain with bounded rather than marginal 
fluctuations. Also, the investigation of ordered cases that do not
stem from a substitution rule might add some insight as they can
possess a higher degree of complexity. Unfortunately, renormalisation
techniques will probably be unsuitable here.

Of primary interest are, of course, extensions to higher dimensions. 
A natural first step is to study two-dimensional statistical systems with 
two-dimensional rather than one-dimensional variations of the 
interactions. As can be seen from some special cases \cite{BGB},
analytic results are possible in principle, but they are not generic
and solvable cases might actually be misleading. Further analysis is
necessary, and some results can be expected from numerical approaches,
especially since structures generated 
by substitution rules are well-suited for (numerical) renormalisation. 
First results in this direction have been obtained recently \cite{S}
and will be extended soon.

\section*{Acknowledgements}
Part of this work was done while JH was visiting Melbourne University. 
It is a great
pleasure to thank Paul Pearce for his kind hospitality. Financial support 
from the German Science Foundation (DFG) and the Studienstiftung 
des Deutschen Volkes is gratefully acknowledged.

\setcounter{section}{0}
\renewcommand{\thesection}{Appendix \Alph{section}:}
\renewcommand{\thesubsection}{\Alph{section}.\arabic{subsection}}
\renewcommand{\theequation}{\Alph{section}.\arabic{equation}}

\section{Solution of the eigenvalue problem for two-letter substitutions}
\setcounter{equation}{0}

We consider two-letter chains generated by substitution rules
\begin{equation}
\varrho:\; \begin{array}{ccc} a&\to&a w_a\\b&\to&b w_b\end{array}
\end{equation}
which have $\lambda_{2}=1$ as the second-largest eigenvalue 
of the corresponding substitution matrix. As a consequence, one finds 
the constraints
\begin{equation}\label{eq:wawb}
p_a = \frac{\#_a(w_a)}{|w_a|} = \frac{\#_a(w_b)}{|w_b|} \; , \qquad
p_b = \frac{\#_b(w_a)}{|w_a|} = \frac{\#_b(w_b)}{|w_b|} \; ,
\end{equation}
on the words $w_a$ and $w_b$ and the letter frequencies $p_{a,b}$ 
in the limit word,
where $\#_{\alpha}(w_{\beta})$ denotes the number of letters $\alpha$ 
contained in the word $w_{\beta}$. We parametrise the criticality condition 
(\ref{eq:critcond}) $\varepsilon_a^{p_a}\varepsilon_b^{p_b}=1$
(setting $h=1$) for the corresponding Ising chain by
\begin{equation}
\varepsilon_a = r^{-l/2} \; , \quad \varepsilon_b = r^{k/2}\; ,
\end{equation}
with {\em coprime}\/ integers $k$ and $l$, 
such that $l p_a = k p_b$. Then $k+l$ divides $|w_a|$ and $|w_b|$.

The scaling exponent of the critical spectrum is a simple 
expression of the leading eigenvalues of two matrices
\begin{equation}
\bm{M}^\pm = \left(\begin{array}{@{\,}cc@{\,}} P^\pm_{aa} & P^\pm_{ba}\\ 
                                               P^\pm_{ab} & P^\pm_{bb}
             \end{array}\right)
\end{equation}
where the entries $P^\pm_{xy}$ are defined in 
Eqs.~(\ref{pminus}) and (\ref{pplus}). They are polynomials in $r$ and 
$r^{-1}$. In what follows, we prove that $\bm{M}^\pm$ have 
an eigenvalue $\lambda_2 =1$, {\em independently}\/ of the value 
of $r$ and of the detailed form of the substitution rule $\varrho$. 
Since we have $\bm{M}^+(r) = \bm{M}^-(r^{-1})$, 
it suffices to concentrate on $\bm{M}^+$.

We define two polynomials $P^{a,b}$ corresponding to the words 
$w_a$ and $w_b$ by
\begin{equation}\label{eq:pab}
P^a(r,r^{-1},u)  :=  (P^+_{aa}-1) + u P^+_{ab} \; ,\qquad
P^b(r,r^{-1},u)  :=  P^+_{ba} + u (P^+_{bb}-1) \; .
\end{equation}
Taking the frequencies $p_{a,b}$ of the letters as given, the possible words
$w_a$ and $w_b$ are subject to the constraints (\ref{eq:wawb}).
We define an {\em elementary polynomial}\/ 
$P^e(r,r^{-1},u)$ to be the polynomial $P^{a,b}$ corresponding to a 
word $w_e$ of minimal length fulfilling this condition. 
Since $k$ and $l$ are coprime, each $w_e$ consists of $k$ $a$'s and $l$
$b$'s and leads to one of ${k+l \choose l}$ different elementary 
polynomials possible. 

\begin{figure}[ht]
\centerline{\epsfysize=80mm  \epsfbox{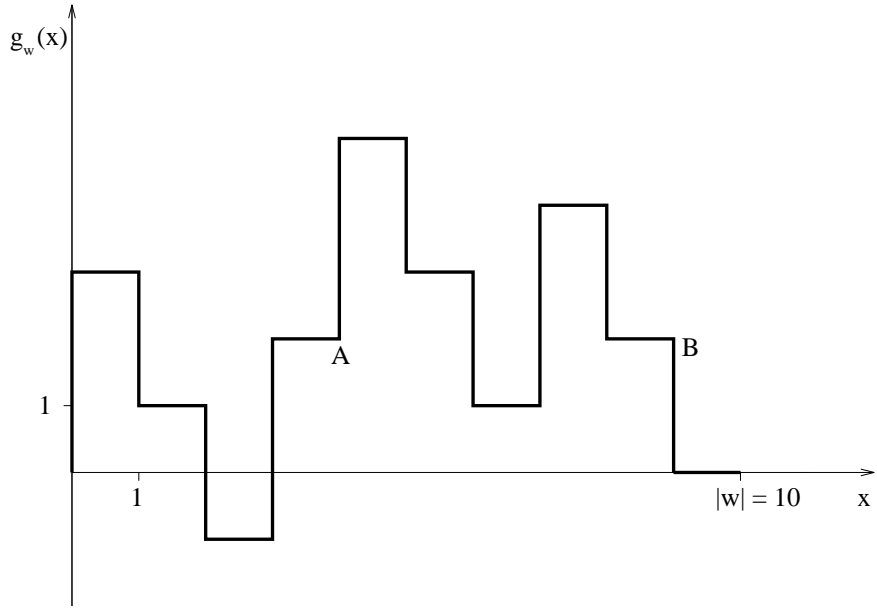} }
\caption{\small Step function $g_w(x)$ 
(\protect\ref{eq:gw}) of the word $w=abbaabbabb$. The
endpoints of an elementary polynomial that can be separated are
marked by `A' and `B'.\label{fig1}}
\label{step}
\end{figure}

Let in the following $w$ be $w_a$ or $w_b$.
We now introduce a graphical representation of the word $w$ and the
accompanying polynomials. For given $w$, construct iteratively a step
function $g_w: [0;|w|] \to \mathbb{Z}$ as follows:
\begin{eqnarray}
g_w(0)&=&0 \nonumber \\ \label{eq:gw}
g_w(x)&=& \left\{ \begin{array}{@{\,}l@{\quad}r@{\,}} 
g_w(k-1)+l & \mbox{if $w^k=a$} 
\\ g_w(k-1)-k& \mbox{if $w^k=b$}\end{array}\right\}\quad 
\mbox{for $k-1<x\le k\;$.}
\end{eqnarray}
Note that the criticality conditions imply $g_w(|w|)=0$. An example for
the case $k=2$, $l=3$ is shown in Figure~\ref{fig1}.
For $w=w_a$ ($w=w_b$), $P^+_{aa}-1$ (resp.\ $P^+_{ba}$) is given by the 
``sum over the {\em upward}\/ steps'' of $g_w$ (proceeding in positive 
$x$-direction), while $P^+_{ab}$ (resp.\ $P^+_{bb}-1$) 
is given by the ``sum over
the {\em downward}\/ steps'', each step contributing
a term $r^{g_w(k)}$. In our example (for $w=w_a$),  
$P^+_{aa}-1 = r^3+r^2+r^5+r^4$ and $P^+_{ab} = r+r^{-1}+r^3+r+r^2+1$.

In a first step, we now show that $P^{a,b}$ can be expressed as sums
over elementary polynomials $P^{e}$.

\noindent
{\em Lemma:}\quad For an arbitrary word $w$ fulfilling the criticality
condition for given $p_{a,b}$, there are integers $a_i$ and elementary
polynomials $P^e_i$, such that
\begin{equation}\label{lem}
P^{a,b}(r,r^{-1},u) = \sum_i r^{a_i} P^e_i(r,r^{-1},u)
\end{equation}

\noindent
{\em Proof:}\quad Consider the substrings $s_j$ of $w$ of length $k+l$
starting with the letter $w^j$. For each $s_j$ containing more than
$k$ $a$'s, there has to be an $s_i$ with less than $k$ $a$'s, because
of the criticality condition for $w$. Since the number of $a$'s contained in
successive substrings $s_j$ and $s_{j+1}$ differs at most by one,
there is at least one substring consisting of $k$ $a$'s and $l$
$b$'s. In the graphical representation, this appears as a restriction of 
$g_w$ to an interval of length $k+l$ with the endpoints of the 
graph taking the same value.
The corresponding term within $P^{a,b}$ is obviously some elementary 
polynomial times a 
power of $r$. We may now eliminate this substring from $w$ 
(cut out the interval in the graphical picture) and start the argument again.

We proceed by showing that the elementary polynomials
differ only in a (polynomial) factor that is independent of $u$.

\noindent
{\em Proposition:}\quad Given $k$ and $l$, there are polynomials 
$P_k(r,r^{-1})$ and $P_l(r,r^{-1})$ such that for every elementary 
polynomial $P^e$
\begin{equation}\label{prop}    
P^e(r,r^{-1},u) = Q^e(r,r^{-1}) \left[P_k(r,r^{-1}) + u P_l(r,r^{-1})\right]
\end{equation}
where $Q^e(r,r^{-1})$ is a polynomial that does not depend on $u$.

\noindent
{\em Proof:}\quad An elementary polynomial 
$P^e$ takes the following general form
\begin{equation}
P^e = r^{l-c_1k} + r^{2l-c_2k} + \ldots + r^{kl-c_kk} + u
\left(\sum_{j=1}^{c_1} r^{-jk} + \sum_{j=c_1}^{c_2} r^{l-jk} + \ldots
+ \sum_{j=c_k}^{l} r^{kl-jk}\right)
\end{equation}
where $0\le c_1\le c_2 \le \ldots \le c_k \le l$ are integers. Defining
\begin{equation}
P_k := \sum_{j=0}^{k-1} r^j = \frac{1-r^k}{1-r}\; ,\quad
P_l := \sum_{j=1}^{l} r^{-j} = \frac{r^{-l}-1}{1-r}\; ,
\end{equation}
the proof then follows by direct calculation. 
We remark that with this choice of
$P_k$ and $P_l$, $Q^e$ can also be shown to be a polynomial in $r$ and
$r^{-1}$. 

Now, we are in the position to complete our argument.
{}From Eqs.~(\ref{eq:pab}), (\ref{lem}) and (\ref{prop}), we conclude that
\begin{equation}
\frac{P_{aa}-1}{P_{ab}} = \frac{P_k}{P_l} = \frac{P_{ba}}{P_{bb}-1}\; .
\end{equation}
Thus it follows that $\det(\bm{M}^+ - \mathbb{I}) =0$, 
and hence $\lambda_2 = 1$ is an eigenvalue of $\bm{M}^+$ for an
{\em arbitrary}\/ value of $r$. The leading eigenvalues of $\bm{M}^{\pm}$
are given by Eq.~(\ref{mu}).

\section{Renormalisation formalism for general 
$\protect\bm{n}$-letter substitution rules}
\setcounter{equation}{0}

In the above discussion, the coupling constants were chosen according to
a substitution rule of the special form (\ref{subs}). While any (infinite) 
substitution chain with two letters can be generated that way, this is
no longer the case for three or more different letters (resp. couplings). 
In this Appendix, we present a generalised version of the renormalisation 
formalism to deal with the general case. Consider a substitution rule 
of the form
\begin{equation} \label{subs2}
\varrho:\;\; a_i \to w_i = w_{i1} a_i w_{i2} \; .
\end{equation} 
The main problem in the general case is to maintain the recursive structure
of the renormalisation procedure within the elimination process. By fixing 
the first letter of the substitutes in Eq.~(\ref{subs}), the appropriate 
couplings of the renormalised chain were given. 
This is no longer possible here; however, we can proceed as follows. 
\begin{itemize}
\item
Fix any letter $a_i$ within the 
word $w_i = \varrho(a_i)$, thus defining two words $w_{i1}$ and $w_{i2}$ 
as shown in Eq.~(\ref{subs2}). Note that the letter $a_i$ appears
in the word $w_i$ (at least for a suitably chosen power $\varrho^k$ of
the substitution rule) for primitive substitutions. 
\item
Redefine the substitution rule as a function on {\em pairs}\/ of letters 
in the following way
\begin{equation}\label{subs3}
\varrho \leadsto \bar{\varrho}:\;\; a_i (a_j) \to a_i w_{i2} w_{j1} (a_j)
\end{equation}
Obviously, (\ref{subs3}) leads to the same limit chain as (\ref{subs2}). 
The substitution matrix of $\bar{\varrho}$ has a dimension of at most $n^2$
(pairs $a_ia_j$ that do not appear in the original chain need not be 
included in $\bar{\varrho}$). Actually, $\bm{M}_{\bar{\varrho}}$ 
is just the substitution matrix of the corresponding {\em site}\/-problem, 
where the type of site follows the sequence, and the bond (resp.\ the 
interaction along it) is a function of its two endpoints \cite{MQ,TBB}. 
Note, however, that the couplings here are still attached to the bonds.
\item
In the renormalisation formalism, now define $S$-matrices, fields and 
asymmetry-parameters corresponding to pairs of letters
\begin{equation}
S_{a_i|a_j} = \left(
\begin{array}{@{\;}cc@{\;}} \varepsilon_{a_i}^{-1} \kappa_{a_ia_j}^+ \Lambda
& -\varepsilon_{a_i}^{-1} h_{a_ia_j}\\ -\varepsilon_{a_j}^{-1}h_{a_ia_j}& 
\varepsilon_{a_j}^{-1} \kappa_{a_ia_j}^- \Lambda \end{array} \right) \; .
\end{equation}
Renormalisation again means inverting the substitution process by 
integrating out all degrees of freedom attached to the words $w_{i1}$ 
and $w_{i2}$. We obtain renormalised $S$-matrices through
\begin{equation}
\tilde{S}_{a_i|a_j} = S_{a_i|w_{i2}^1} \ast S_{w_{i2}^1|w_{i2}^2} \ast\ldots 
\ast S_{w_{i2}^{|w_{i2}|}|w_{j1}^1} \ast\ldots\ast S_{w_{j1}^{|w_{j1}|}|a_j}
\; .
\end{equation}
\item
The RT (and their linear orders in $\Lambda$) for the
(at most) $3n^2$ parameters are then obtained analogous to Section~3.
\end{itemize}
For a classification of the critical behaviour based on the 
fluctuations we need to know about the spectrum 
$\sigma_{\bar{\varrho}}$ of $\bm{M}_{\bar{\varrho}}$ in 
dependence of the spectrum $\sigma_\varrho$ of $\bm{M}_\varrho$. 
For an appropriate power of $\varrho$, it is indeed possible to show 
that $\sigma_{\bar{\varrho}}$ contains $\sigma_\varrho$ and that all 
additional eigenvalues are either 1 or 0.
We only sketch the proof here. 
Without restriction of the general case, we can assume 
that the first and the last letters of all words $w_i$ 
remain fixed under the substitution; 
that is $w_i^1 = [\varrho(w_i)]^1$ for the first letter, 
and equivalently for the last letter. This is always fulfilled 
by a finite power $\varrho^r$ of the substitution rule. 
We assume the $n$ words $w_i$ to have $k$ different first letters 
and $j$ different last letters. Taking a look at the associated 
pair substitution matrix $\bm{M}_{\bar{\varrho}}$, it is easy to show that
\begin{equation}
\mbox{tr} (\bm{M}_{\bar{\varrho}}) = \mbox{tr}(\bm{M}_{\varrho})
+ n + kj - k - j \; .
\end{equation}
Since we always have $\bm{M}_{\bar{\varrho}^m} = 
(\bm{M}_{\bar{\varrho}})^m$ and the numbers $k$ 
and $j$ remain fixed because of the above assumption, the traces of 
$\bm{M}_{\bar{\varrho}^m}$ and $\bm{M}_{\varrho^m}$ differ only 
by the constant $n+kj-k-j$ for arbitrary $m$. We 
conclude that $\sigma_{\bar{\varrho}}$ containes $\sigma_\varrho$ and 
the all additional eigenvalues are fixed points under any power, 
thus $1$ or $0$.

We will, however, not give a detailed discussion of the general case here, 
but rather illustrate the method by applying it to a special example. 
Consider the substitution rule of the ``circle sequence''. This example of a 
quasiperiodic chain (in the sense that it has a pure point Fourier spectrum) 
has been studied numerically in Ref.~\cite{L}
\begin{equation}
\varrho:\;\; \begin{array}{rclcl}
a &\to& cac& \to& \underline{a}bcaccacabcac \\
b &\to& accac&\to& caca\underline{b}cacabcaccacabcac \\
c &\to& abcac&\to& \underline{c}acaccacaabcaccacabcac 
\end{array}
\end{equation}
with $\varrho^2$ being of the form (\ref{subs3}).
The eigenvalues of $\bm{M}_{\varrho^2}$ are $\tau^6$, $1$, and $\tau^{-6}$, 
where $\tau=(1+\sqrt{5})/2$ is the golden mean -- we thus expect marginal 
fluctuations. The asymptotic frequencies of the letters are 
$p_a = 2-\tau$, $p_b = \tau -3/2$, and $p_c = 1/2$, respectively. We choose
$\varepsilon_a = rs^{-\tau}$, $\varepsilon_b = rs$, and 
$\varepsilon_c =r^{-1}s$ as the parametrisation of the critical couplings
$\varepsilon_a^{p_a} \varepsilon_b^{p_b} \varepsilon_c^{p_c}=1$.
Since only the five pairs $ab$, $ac$, $bc$, $ca$, and $cc$ appear in the 
chain, the dimension of the pair transfer matrix $\bm{M}_{\bar{\varrho}^2}$ 
is just five. 
The spectrum $\sigma_{\bar{\varrho}^2}$ consists of the spectrum
$\sigma_{\varrho^2}$ and of two additional eigenvalues 1. 
Identifying pairs of couplings with the first one, we clearly have
$p_{ab}+p_{ac}=p_a$, $p_{bc}=p_{b}$, and $p_{ca}+p_{cc}=p_c$. 
Thus the vector \bm{y} of the logarithms of the reduced couplings, 
$y_{xy} = 2 \log (h_{xy}/\varepsilon_x)$, 
is sill perpendicular to the Perron-Frobenius eigenvector of 
$\bm{M}_{\bar{\varrho}}$ and converges in the renormalisation limit to a 
linear combination of two eigenvectors \bm{v^{1,2}} corresponding to 
an eigenvalue 1 of $\bm{M}_{\bar{\varrho}}^t$: 
$\bm{y}\to\log r \,\bm{v^1} + \log s \,\bm{v^2}$. 
While both parameters, $r$ and $s$, enter the transformation matrices 
$\bm{M}^\pm$ of $\bm{\Lambda}_\pm$, 
$s$ can actually be eliminated by a similarity transformation. 
Such a behaviour is always to be expected 
if a parameter is connected to an eigenvector (like \bm{v^2} in this
case) that corresponds to an {\em additional}\/ eigenvalue 1 of 
$\bm{M}_{\bar{\varrho}^2}^t$, not contained in $\sigma_{\varrho^2}$. 
The Perron-Frobenius eigenvalues of $\bm{M}^\pm$ can be determined 
explicitly, and we obtain for the scaling exponent
\begin{equation}
z = \frac{\log\left(5 + 2(r+r^{-1}) +5 + 
         2\sqrt{(r+r^{-1}+2)^2+(r+r^{-1}+2)}\right)}{6 \log(\tau)}\; .
\end{equation} 
Finally, we like to show an example of a three-letter
substitution rule 
\begin{equation}
\varrho:\;\;
 \begin{array}{ccl} a& \to& bca\\ b&\to &bcb\\ c&\to &ca \end{array}
\end{equation}
with bounded fluctuations, 
which nevertheless has $\lambda_2 = 1$ as the second-largest 
eigenvalue of the substitution matrix.
This is the bond substitution that corresponds to the site-problem 
given by $\varrho_s: a \to aba\, ;\; b \to ba$.  
Fluctuations are indeed bounded here, since we just obtain the Fibonacci 
chain by identifying $a$ and $b$, but also by identifying $a$ and $c$. 
Calculating the scaling exponent by the above method, we obtain $z=1$, 
as was to be expected.

\end{document}